\documentclass[twocolumn,prd,aps,superscriptaddress
,showpacs,
nofootinbib,preprintnumbers,floats,floatfix]{revtex4}

\usepackage{color}
\usepackage{amsmath}
\usepackage{amssymb}
\usepackage{latexsym}
\usepackage{graphicx}
\usepackage{hyperref}
\usepackage{bm}

\newcommand{\fnl}{\ensuremath{f_{\mathrm{NL}}}}

\newcommand{\mpl}{m_{\rm Pl}}

\begin{document}

\title{Curvaton scenarios with inflaton decays into curvatons}
\author{Christian T.~Byrnes}
\affiliation{Astronomy Centre, University of Sussex, Falmer,  
Brighton BN1 9QH, UK}
\author{Marina Cort\^es} 
\affiliation{Institute for Astronomy, University of Edinburgh, Blackford Hill, Edinburgh EH9 3HJ, UK} 
\affiliation{Centro de Astronomia e Astrof\'isica da Universidade de Lisboa, Faculdade de Ci\^encias, Edif\'icio C8, Campo Grande, 1769-016 Lisboa, Portugal}
\affiliation{Perimeter Institute for Theoretical Physics,
31 Caroline Street North, Waterloo, Ontario N2J 2Y5, Canada}
\author{Andrew R.~Liddle} 
\affiliation{Institute for Astronomy, University of Edinburgh, Blackford Hill, Edinburgh EH9 3HJ, UK} 
\affiliation{Perimeter Institute for Theoretical Physics,
31 Caroline Street North, Waterloo, Ontario N2J 2Y5, Canada}
\date{\today}
\begin{abstract}
We consider the possible decay of the inflaton into curvaton particles during reheating and analyse its effect on curvaton scenarios. Typical decay curvatons are initially relativistic then become non-relativistic, changing the background history of the Universe. We show that this change to the background is the {\em only} way in which observational predictions of the scenario are modified. Moreover, once the required amplitude of perturbations is fixed by observation there are no signatures of such decays in other cosmological observables. The decay curvatons can prevent the Universe from becoming dominated by the curvaton condensate, making it impossible to match observations in parts of parameter space. This constrains the branching ratio of the inflaton to curvaton to be less than of order $0.1$ typically. If the branching ratio is below about $10^{-4}$ it has negligible impact on the model parameter space and can be ignored. 
\end{abstract}

\pacs{98.80.Cq}

\maketitle

\section{Introduction}

There is increasingly strong evidence that observed structures are sourced by initially super-horizon adiabatic perturbations \cite{PlanckXIII,PlanckXX,Ade:2015ava}. However, the same observations leave open the question of whether those perturbations  were acquired during inflation by the same degree of freedom whose energy density causes inflation, or by an additional degree of freedom such as a light scalar field. Either mechanism is capable of generating super-horizon purely adiabatic perturbations. A simple example of the latter is the curvaton model \cite{seminal} (see Ref.~\cite{molllind} for earlier related ideas), and hence there is considerable interest in assessing how viable the curvaton model is against observations. This has been the topic of a number of recent papers \cite{kobayashi,Enq-Tak,BCL,Hardwick:2015tma,Vcurvatonis,Vcurvatonis2}. These have tested a wide regime of model space, including cases where the observed perturbations come from a mixture of inflaton and curvaton perturbations that feed into the perturbed thermalized Universe after each field has decayed \cite{Langlois:2004nn}. The key quantity is the relative amount of inflaton-like and curvaton-like perturbations represented in this final thermal bath, the thermalization guaranteeing that the perturbations have become of purely adiabatic form.

It is likely that the inflaton decays also into curvaton particles, at least at some level  \cite{LM,Sasaki:2006kq,supercurv}. A plausible {\em lower} limit comes from Planck-suppressed gravitational decay interactions, through which the inflaton decays into all available particle species at the same rate \cite{Watanabe:2007tf}. There may also be direct decay of inflatons to curvatons via interaction terms in the action such as $\phi \sigma^2$ and $\phi^2\sigma^2$, as are commonly considered in reheating models. Further, since the curvaton must be able to decay into the thermal bath, it must be possible for curvatons to be created by the inverse process from any thermal bath generated by inflaton decays. Hence we should consider it to be more or less guaranteed that some fraction of the inflaton energy density will find its way into curvaton particles. Investigating the consequences of this inflaton-into-curvaton decay is the purpose of this article. We will for the most part adopt a phenomenological approach where we simply dictate that some fraction $f$ of the inflaton energy density is transmuted to curvatons at the time of inflaton decay. For a consideration of non-perturbative curvaton decay, see e.g.\ Ref.~\cite{nonpert}.

Our principal goal is to assess the effect of adding such a term to curvaton scenarios, which we do within the context of the simplest curvaton model \cite{BL} through extension of the analysis we carried out in Ref.~\cite{BCL}. As we will see, this model continues to be able to fit all available observations. In its original form the model contains two massive but otherwise non-interacting fields during inflation; it is specified by five parameters (two masses, two field decay rates, and the early-time value of the curvaton field). To this we add the branching ratio $f$.

As the curvaton condensate may have orders of magnitude less energy density than the inflaton when the inflaton decays, it is readily possible for the decay curvatons to swamp the energy density of the curvaton condensate. One might expect this to have a radical effect on predictions from the curvaton scenario. In particular it has been argued that the consideration of this decay, and of the presence of the resulting curvaton particles, could alter the predicted non-gaussianity  \cite{LM}, potentially rescuing curvaton models with large $\fnl$ that are otherwise ruled out by {\it Planck} observations (as described in Ref.~\cite{Ade:2015ava}, particularly Sec.~11.2). However, we will show here that this is not the case. The subsequent decay of the produced curvatons into thermal radiation, at the time of curvaton decay, transmits the original inflaton perturbations into the final thermal bath, and as they are adiabatic super-horizon perturbations throughout the process, their ultimate contribution to the perturbations is unchanged. This is simply invoking the same argument that we can compute the final adiabatic perturbations from single-field inflation independent of the details of any intermediate reheating process. Here we have initial perturbations from the inflaton that are adiabatic, and from the curvaton which are isocurvature, but the only thing which matters is the relative proportions of the two contributions to the final thermal bath which locks in the perturbations as adiabatic. We do not consider the case that isocurvature perturbations persist after curvaton decay, beyond big bang nucleosynthesis, and potentially up to the present \cite{Smith:2015bln}.

Although the inflaton perturbations are unaffected by possible routing via curvaton channels, there does remain a physical effect that can constrain both the curvaton  scenario and the possible values of the corresponding branching ratio. The effect is that the introduction of decays to curvatons may change the overall background history of the Universe. That is, if the decay curvatons come to dominate the energy density of the background, while having a different evolution law from the other decay products of the inflaton, there will be observable consequences for the model. This would change the relative mix of inflaton-like and curvaton-like perturbations in the final thermal bath, altering the observational predictions. However, as we will show even this effect is mitigated, because having fixed the magnitude of inflaton perturbations we must then tune the curvaton decay rate so that the overall observed total perturbation amplitude is generated. Hence if the background evolution is changed, we must simultaneously modify the curvaton decay timescale to ensure that the contribution of the curvaton perturbations to the total is maintained. This then ensures that other observational quantities are unchanged, i.e.\ claims in the literature that the predicted non-gaussianity may change \cite{LM,Sasaki:2006kq,supercurv} are not realised once the perturbation amplitude is fixed.

Nevertheless, it may be that the changed background evolution due to decay curvatons makes it impossible to generate the required mix of inflaton and curvaton perturbations, particularly in what would have been the curvaton-dominated limit. This does give a significant constraint on the branching ratio. If the curvaton dominated limit is not possible, then the corresponding value of $\fnl=-5/4$, which one may consider the preferred value for the quadratic curvaton scenario, cannot be reached and instead $\fnl>-5/4$.

In the next section we provide the key equations defining the observables and provide qualitative statements regarding the effect of introducing the decay fraction $f$. In Sec.~\ref{sec:observation} we show how concrete constraints on the simplest curvaton scenario depend on $f$, before concluding in Sec.~\ref{sec:conclusions}. Note that we use the non-reduced Planck mass throughout this paper.

\section{Evolution and perturbations}

We follow the notation and methods of our previous paper \cite{BCL}, and here just summarize the main equations. The inflaton field is denoted $\phi$ and the curvaton field $\sigma$, with potential
\begin{equation}
V(\phi,\sigma) = \frac{1}{2}m_\phi^2 \phi^2+ \frac{1}{2}m_\sigma^2 \sigma^2 \,.
\end{equation}
The number of $e$-foldings of inflation from field values $\phi$ and $\sigma$ is given by 
\begin{equation}
N  = 2\pi \frac{\phi^2 + \sigma^2}{m_{\rm Pl}^2}\,,
\end{equation}
where $m_{\rm Pl}$ is the (non-reduced) Planck mass. We have neglected the small contributions from the field values at the end of inflation, and in practice the $\sigma$ contribution will always be negligible under our assumptions. We allow both inflaton and curvaton perturbations to contribute to the  total power spectrum, which is guaranteed to be solely adiabatic once both fields have decayed into the  thermal bath. In this article we make the simplifying assumption that the curvaton does not drive a period of inflation \cite{Dimopoulos:2011gb}, which reduces the number of cases that need to be considered.

As we remarked in the introduction, to fully specify the model outcome we need to know both the initial conditions and the way in which the fields decay into the late-time thermal bath. We need only consider the initial value of the curvaton field $\sigma_*$, not the inflaton field due to the usual inflationary attractor behaviour that renders predictions independent of the initial value of the inflaton field set well before observable scales left the horizon. We can phenomenologically represent the decay into the thermal bath via decay rates $\Gamma_\phi$ and $\Gamma_\sigma$. Hence the model so far possesses five adjustable parameters, which was the model space investigated in our earlier paper \cite{BCL}. This paper will add a branching ratio $f$ for inflaton-to-curvaton decays.

We recap the main equations for the observables from Ref.~\cite{BCL}. The total power spectrum is made up of uncorrelated inflaton and curvaton parts
\begin{equation}\label{ps}
P^{\rm total}_{\zeta}= P^{\phi}_{\zeta}+P^{\sigma}_{\zeta}\,.
\end{equation}
The inflaton contribution is the usual
\begin{equation}
P_\zeta^{\phi} = \frac{4 m_\phi^2 \phi_*^2}{3 m_{\rm Pl}^4} \,2N_* \,,
\end{equation}
where * refers to the parameter value when observable scales crossed the Hubble radius during inflation. Once parameter values are chosen the relevant scale can be computed, fixing $\phi_*$ and $N_*$, as shown in Ref.~\cite{BCL} and subsequently explored in detail in Refs.~\cite{Vcurvatonis,Vcurvatonis2}. Hence $m_\phi$ alone is required to fix the inflaton contribution to the perturbations, due to the usual conservation of super-horizon scale adiabatic perturbations.

The curvaton contribution is \cite{Lyth:2002my} 
\begin{equation}\label{Psigma}
P_\zeta^{\sigma} = \frac{8r_{\rm dec}^2}{27 \pi}\frac{m_\phi^2 \phi_*^2}{\sigma_*^2 m_{\rm Pl}^2}\,.
\end{equation}
where $r_{\rm dec}$ is a measure of the ratio of curvaton to background density at the time the curvaton decays into the thermal bath. We will define this properly later. Importantly, we will see that all the equations of this section remain valid in the presence of inflaton-to-curvaton decays provided $r_{\rm dec}$ is suitably generalized from the expression in Ref.~\cite{BCL}.

The observed perturbation amplitude is \cite{WMAP9, PlanckXIII}
\begin{equation} \label{ampNorm}
P_\zeta^{\rm obs} \sim 2.2\times 10^{-9}\,,
\end{equation}
and parameter values need to be chosen to reproduce it. Once $m_\phi$ is specified the inflaton's share of the total perturbations is fixed (apart from a mild dependence on $N_*$ \cite{BCL,Hardwick:2016whe}, computed as in Ref.~\cite{LiddleLeach}), and if we consider $\sigma_*$ as a further free parameter, the amplitude condition fixes $r_{\rm dec}$. Models are viable only if the required $r_{\rm dec}$ is achievable, otherwise the correct amplitude of perturbations cannot be generated. Typically we shall display model constraints considering $m_\phi$, $m_\sigma$, and $\sigma_*$ as free parameters.

The expressions for the model observables --- the spectral index $n_{\rm S}$, tensor-to-scalar ratio $r$, and non-Gaussianity parameter $\fnl$ ---  are as in our previous paper. The slow-roll parameters are defined by
\begin{equation}
\epsilon=-\frac{\dot{H}}{H^2}\simeq \frac12 \left(\frac{V_{\phi}}{3H^2}\right)^2, \;\;\eta_\phi = \frac{V_{\phi \phi}}{3H^2},\;\;\eta_\sigma = \frac{V_{\sigma \sigma}}{3 H^2},
\end{equation}
and the observables are, in the usual notation, \cite{Wands:2002bn,Lyth:2002my,Sasaki:2006kq,Enqvist:2009eq,Enq-Tak,BCL}
\begin{eqnarray}\label{ns}
n_{\rm S}-1 & = & \left(1-\frac{m_{\phi}^2}{m_{\rm single}^2}\right) (-2\epsilon+2\eta_{\sigma}) \nonumber  \\
&& \qquad +\frac{m_{\phi}^2}{m_{\rm single}^2}\left(-6\epsilon+2 \eta_{\phi}\right), \\
\label{r_tens}
r & = &16\, \epsilon \frac{m_{\phi}^2}{m_{\rm single}^2} \,,\\
\label{fnl}
f_{\rm NL} &= & \frac{5}{12} \left(1-\frac{m_{\phi}^2}{m_{\rm single}^2}\right)^2 \times \\ \nonumber
 & & \quad \left(\frac{3}{r_{\rm dec}}+2(\beta-6)-2 r_{\rm dec}(\beta-3) \right)\,. \quad \quad
\end{eqnarray}
Here $m_{\rm single}\simeq1.4\times10^{-6}\mpl$ is the value that the inflaton mass would have if the perturbations were to be purely inflaton-generated, and $\beta=3(1+w)$ measures the rate of redshifting of the inflaton decay products at the time the curvaton decays, where $w$ is the overall equation of state of this component (or components). For the most commonly considered case of pure radiation $\beta = 4$, while $\beta = 3$ for pure non-relativistic matter, and is intermediate between those two for a two-component fluid such as we will be considering.

The $\fnl$ formula merits some comment. In our formalism it follows straightforwardly from attributing the decay curvatons to the inflaton-originated contribution, through the generalized $r_{\rm dec}$ defined below. In the limit of negligible inflaton perturbations, $m_\phi \ll m_{\rm single}$, the formula was given in this form in Ref.~\cite{Enqvist:2009eq}, then generalized to include the inflaton contribution in Ref.~\cite{Enq-Tak}. We are further generalizing it via including decay curvatons into the definition of $r_{\rm dec}$, which does not change the form of this equation. However, in a different notation and physical viewpoint, a suitable expression was first given by Sasaki et al.~\cite{Sasaki:2006kq}. In their calculation the decay curvatons were considered to give a short-scale contribution to the variance of the curvaton field, and were considered as part of the curvaton field, albeit with a homogeneous distribution. Their leading term, inversely proportional to $r_{\rm dec}$, matches ours; their definition of $r_{\rm dec}$ differs but the difference is cancelled by the $(1+\Delta_s^2)$ term in their Eq.~(102). Their full expression can be shown to be equivalent to ours (in the  $m_\phi \ll m_{\rm single}$ limit that their calculation imposes) if the inflaton decay products are dominated either by radiation or decay curvatons; marginal differences in the intermediate regime are irrelevant and likely due to use of a `sudden-decay' approximation. Their formula was also used in the {\it Planck} collaboration analysis of the curvaton model \cite{Ade:2015ava}. 

These equations make evident that, {\em at fixed values of the free model parameters $m_\phi$, $m_\sigma$, and $\sigma_*$,} none of the observables will change (with the modest exception of the sub-dominant terms in $\fnl$, which are affected by the possible change of the background equation of state). This is because they are all determined by the values of just those parameters, plus the known overall amplitude of perturbations which fixes $r_{\rm dec}$. Hence the effect of any new physics, such as inflaton-to-curvaton decays or more generally any change to the background that seeks to modify the fractional density of the curvaton condensate at its decay, proves degenerate with a change in the curvaton decay rate $\Gamma_\sigma$ that restores that fractional density. The degeneracy is enforced by the need to obtain the correct overall amplitude of perturbations.

However, as remarked in the introduction, the inclusion of new physical effects may mean that the necessary value of $r_{\rm dec}$ cannot be realized, hence the model, again at fixed values of $m_\phi$, $m_\sigma$, and $\sigma_*$, may be unable to generate sufficient perturbations. In the original curvaton scenario the curvaton condensate is competing only against radiation and can always completely dominate, $r_{\rm dec} \rightarrow 1$, if the decay happens late enough. But if new physics leads to a long-lived non-relativistic component originating from the inflaton, such as decay curvatons, this gives a maximum achievable $r_{\rm dec}$, which we refer to as $r_{\rm dec}^{\rm max}$, which is less than one. If this maximum is below the value required by observations, fitting the data becomes impossible.

Hence, without carrying out a detailed analysis, we can already conclude
\begin{enumerate}
\item If a model with inflaton-to-curvaton decays can match the observed perturbation amplitude, then all its other predictions are precisely degenerate with a model that has no inflaton-to-curvaton decays, and hence cannot be distinguished from it. 
\item A high level of inflaton-to-curvaton decays may prevent sufficient perturbations being generated for any choice of $\Gamma_\sigma$, preventing the model fitting the data for a given combination of $m_\phi$, $m_\sigma$, and $\sigma_*$. Hence there may be an upper limit on the branching ratio at each location in this space of independent model parameters. 
\end{enumerate}
The remainder of this article makes these qualitative statements quantitative.

\section{Observational constraints}\label{sec:observation}

\subsection{Without inflaton decays to curvaton}

\begin{figure}[t]
\begin{center}
\vspace*{-1.7cm}
\includegraphics[width=0.48 \textwidth]{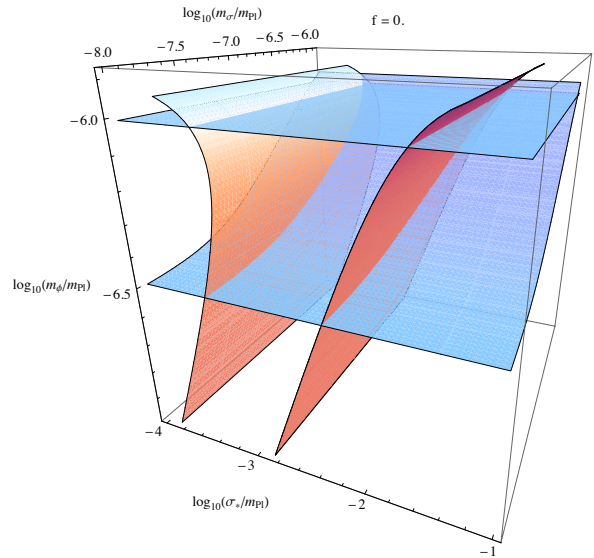}
\vspace*{-2.5cm}
\end{center}
\caption{Observational constraints in the case without inflaton-to-curvaton decays, updating the constraints we presented in Ref.~\cite{BCL}. The upper horizontal plane is the limit from $r$, the bottom near-horizontal surface is from $n_{\rm s}$, the left-hand surface from $\fnl$ and the right-hand surface the limit (via $r_{\rm dec}$ from the observed perturbation amplitude. The allowed region is the portion between these two sets of planes. The highest value of $m_\phi$ shown corresponds to perturbations coming entirely from inflaton, though this is now excluded by the $r$ constraint.}
\label{fzerocheese} 
\end{figure}

We first update the constraints on the standard curvaton scenario, without inflaton-to-curvaton decays, as studied in our earlier paper \cite{BCL}. The main change is that there is now a powerful 95\% upper limit on the tensor-to-scalar ratio coming from the {\it Planck} satellite and BICEP/Keck of $r<0.08$ \cite{BKP}, which comfortably excludes the inflation-dominated regime of the model. By contrast the spectral index constraint is essentially unchanged from before with a 95\% upper limit of about $n_{\rm s} <0.98$ \cite{PlanckXIII,PlanckXX}; both model and dataset-compilation dependence means one cannot reliably add an extra significant digit to this constraint. The spectral index is the main constraint on the curvaton-dominated regime, which has thus been unaffected by observational developments since our previous paper and remains marginally allowed/disallowed depending how strongly one feels about models at 95\% limits. Finally we note a modest tightening of the non-gaussianity constraint to $\fnl < 11$ \cite{Ade:2015ava} (from $\fnl < 14$ previously) that has no qualitative impact on allowed models. Figure~\ref{fzerocheese} shows the updated conclusions. Overall, these observational developments have been more adverse for the inflaton-dominated regime of the model than for the curvaton-dominated regime, and continue to permit a range of models where the perturbations have significant contributions from both sources.

\subsection{Introducing the inflaton to curvaton branching ratio, $f$}

The main consequence of considering the decay of the inflaton via a curvaton channel is the introduction of a new component in the homogeneous background, the decay curvaton particles. The first step is to know where to include the new component when computing $r_{\rm dec}$.  In the simplest curvaton scenario, where there is only radiation when the curvaton condensates decay, this is defined as
$r_{\rm dec} = \rho_{\sigma}/(\rho_{\sigma}+4\rho_{\rm rad}/3)$ where the 4/3 factor is introduced so that 
Eq.~(\ref{Psigma}) holds regardless of which term dominates the denominator. We need to extend this to allow for the decay curvatons.

One might have thought that since these are curvaton particles they should partake in the curvaton component in both the numerator and denominator of $r_{\rm dec}$, hence increasing it. This could in turn decrease the sizeable non-gaussianity which is otherwise typical of curvaton models, and rescue curvaton models by making the typical spectrum more gaussian \cite{LM}. However, as noted above the perturbations of the decayed curvaton particles are super-horizon and remain adiabatic throughout. This means they follow the inflaton spectrum and so the new component produced by the decay of the inflaton is to be included only in the denominator of $r_{\rm dec}$:\footnote{There is no factor 4/3 on the decay curvaton term as we assume they have already become non-relativistic before the curvatons decay. This is sufficient for our calculation of the maximum achievable $r_{\rm dec}$, which corresponds to late curvaton decay. It would need corrected in a scenario where the decay takes place before NR, though the change is anyway tiny compared to other calculational uncertainties.}
\begin{equation}
\label{e:rdec}
r_{\rm dec} = \frac{\rho_{\sigma}}
{\rho_{\sigma}+(4/3)\rho_{\rm rad}+\rho_{\sigma \,{\rm part}}} \,,
\end{equation}
where $\rho_{\sigma \,{\rm part}}$ is the energy density of curvaton particles produced by inflaton decay.
This is the appropriate form for $r_{\rm dec}$ to substitute into the various observational formulae of the previous section.\footnote{Refs.~\cite{Sasaki:2006kq,Ade:2015ava} do include the decay curvatons in the numerator of their decay fraction, but then have to compensate this with an adjustment to the background density via their $\Delta_s^2$.} For the two-component fluid created by inflaton decays, we have 
\begin{equation}
\beta = \frac{4\rho_{\rm rad}+ 3 \rho_{\sigma \,{\rm part}}}{\rho_{\rm rad} + \rho_{\sigma \,{\rm part}}} \,,
\end{equation}
 in the non-gaussianity equation Eq.~(\ref{fnl}).

\begin{figure*}[t]
\begin{center}
\includegraphics[width=0.8 \textwidth]{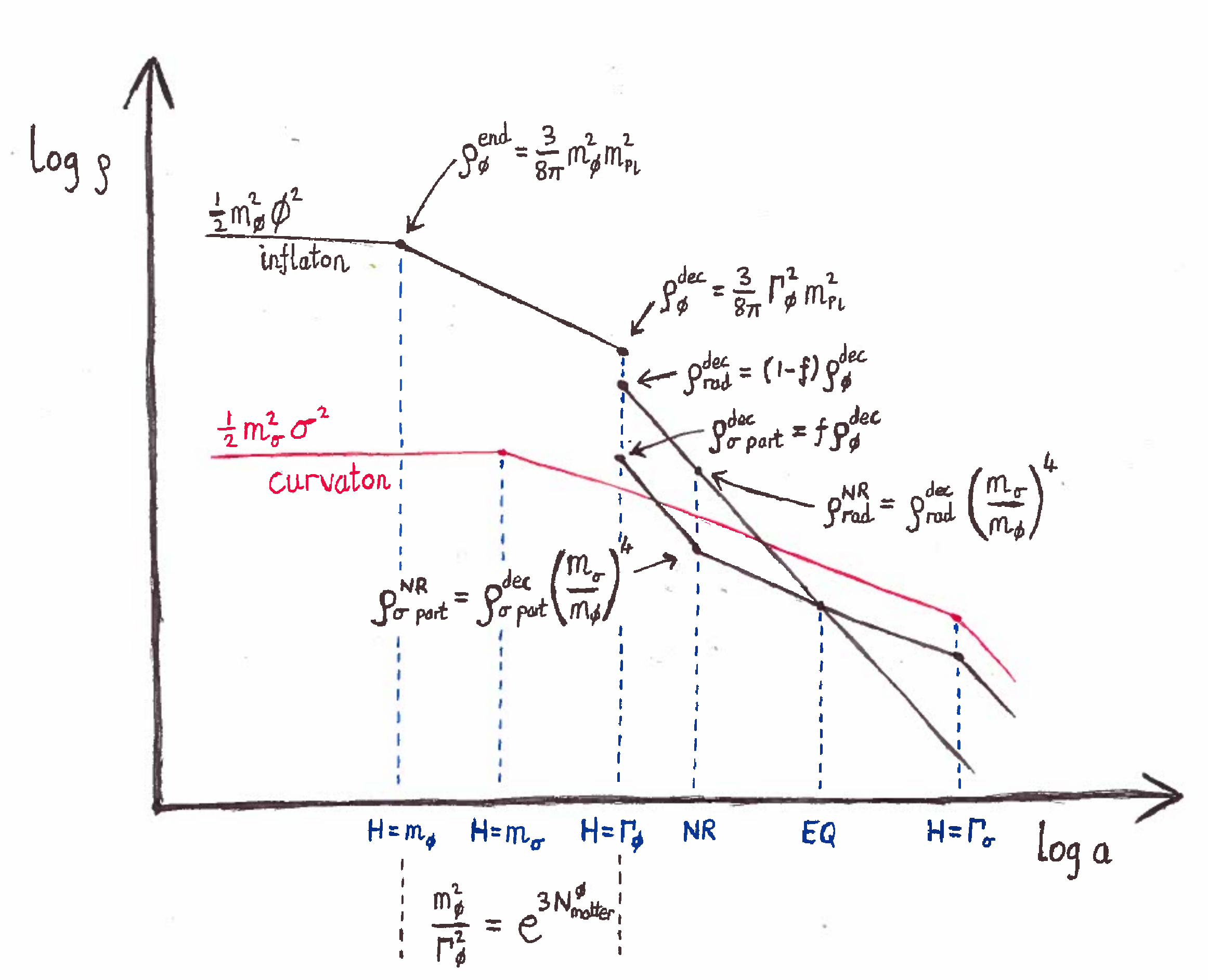}\\ 
\vspace*{-0.5cm}
\end{center}
\caption{A schematic of the main events, showing (on log scales) the evolution of the densities with scale factor. The timing of the various events is mostly given by the Hubble parameter equalling one of the model parameters. From left to right we have the end of inflation, curvaton oscillations beginning, the inflaton decaying, the decay curvatons becoming non-relativistic, the decay curvatons reaching equality with decay radiation, and the decay of the curvatons (both condensate and inflaton-decay generated). The alternative sequence that we also consider puts curvaton oscillations beginning after inflaton decay, though still before NR.}
\label{sequence} 
\end{figure*}

Therefore, if all else is fixed the decay of inflaton to curvaton causes $r_{\rm dec}$ to decrease. However, we still require that $r_{\rm dec}$ takes the value which provides the correct amplitude of perturbations as per Eq.~(\ref{ampNorm}), so the model with reduced $r_{\rm dec}$ does not fit the data. In order to compensate for this effect the curvaton must decay later thus allowing more time for its energy density to increase. Even then there will be a maximum achievable $r_{\rm dec}$, for each $f$, that saturates the energy density of the curvaton condensate at decay. We will call this $r_{\rm dec}^{\rm max}$. If $r_{\rm dec}^{\rm max}$ is not large enough to give the correct perturbation amplitude then the corresponding set of parameters is ruled out. 
Therefore the quantity of interest in estimating the effect of the inflaton to curvaton decay is $r_{\rm dec}^{\rm max}$, since it will determine which models are excluded for each value of the branching ratio. In addition, in cases where the curvaton perturbations dominate the power spectrum, we note that $r_{\rm dec}^{\rm max}$ defines the minimum possible value of $\fnl$. The upper bound on $\fnl$ then places the limit $r_{\rm dec}^{\rm max}>0.19$ at the $95\%$ confidence level \cite{Ade:2015ava}.

\subsection{The sequence of events}

Figure~\ref{sequence} shows the evolution of the various components and the key stages of the evolution. We have restricted the number of cases we need to investigate by assuming that the end of inflaton takes place first (meaning the curvaton is light throughout inflation), curvaton decay takes place last (as we are interested only in calculating the maximum possible $r_{\rm dec}$ as a function of $\Gamma_\sigma$, which occurs in this case), and that there is no period of curvaton-driven inflation. This gives a smaller set of cases than for instance our earlier paper \cite{BCL} and that of Vennin et al.~\cite{Vcurvatonis}, which make only the first of these assumptions. Given those assumptions, there are only two permutations of events that need be considered, which is whether the curvaton becomes massive before or after inflaton decay ($m_\sigma> \Gamma_\phi$ and $m_\sigma < \Gamma_\phi$ respectively). The former is the one that is illustrated in Fig.~\ref{sequence}.

The other reheating scenarios are unlikely to be affected by the introduction of $f$. In the inflating curvaton scenario, all decay curvatons will be quickly washed out during this second period of inflation. If the curvaton decays before the inflaton, it will  never be able to dominate the background energy density, regardless of the value of~$f$.

Including inflaton-to-curvaton decays introduces two new events. Typically the decay curvatons will initially be relativistic, since a characteristic decay route is $\phi+\phi \rightarrow \sigma+ \sigma$ and $m_\phi$ is usually much greater than $m_\sigma$. The decay curvatons then have their momentum redshifted and their density reduces as $1/a^4$. Since the initial total energy per particle is $\sim m_\phi$, the redshifting reduces the kinetic energy to match the rest-mass energy after expansion by a factor $m_\phi/m_\sigma$. We denote this instant as `NR' for the decay curvatons becoming non-relativistic. After this epoch the decay curvatons are losing energy density as $1/a^3$ and catch up with the density of the radiation decay products of the inflaton, doing so after a further expansion by a factor $(1-f)/f$ (being the inverse of the initial relative densities). The epoch of equality of the decay curvatons and radiation is denoted `EQ'.

Note that the curvaton condensate, being composed of zero-momentum particles, necessarily starts to behave as matter-dominated before the decay curvatons, hence `NR' and `EQ' are always to the right of $H=m_\sigma$. Naturally they must also happen after the inflaton decay at $H= \Gamma_\phi$ too. Hence inclusion of these two new phases does not add any extra permutations of events.\footnote{For the purpose of computing $r_{\rm dec}^{\rm max}$, curvaton decay needs to be taken as late as possible and hence indeed does happen last. However in order to reproduce the observed $r_{\rm dec}$ the curvaton will need to decay sometime before $r_{\rm dec}^{\rm max}$ is achieved. This could in fact be earlier in the sequence, for instance before EQ or NR, but this need not concern us.}

We first track the energy density of the inflaton and its decay products through the event sequence.
The energy density at the end of inflation is 
\begin{equation}
\rho_{\phi}^{\rm end}=\frac{3}{8\pi} m_{\phi}^2 m_{\rm Pl}^2 \,.
\end{equation}
If we allow for the inflaton to oscillate for $N_{\rm matter}^{\phi}$ e-foldings after it becomes massive, the energy at inflaton decay will be
\begin{equation}
\rho_{\phi}^{\rm dec}= \rho_{\phi}^{\rm end} \exp^{-3 N_{\rm matter}^{\phi}} = \frac{3}{8\pi} \Gamma_\phi^2 m_{\rm Pl}^2 \,,
\end{equation}
where the last expression follows directly from the expansion being matter-dominated between $H = m_\phi$ and $H = \Gamma_\phi$. At this point the model deviates from the usual reheating scenario because we introduce the inflaton-to-curvaton decay channel. As shown in Fig.~\ref{sequence}, a fraction $f$ of the inflaton's energy density is transferred to curvaton particles, while the radiation decay products are $\rho_{\rm rad}^{\rm dec}=(1-f)\rho_{\phi}^{\rm dec}$. Assuming that $m_{\phi} \gg m_{\sigma}$ then the decay curvaton particles $\rho_{\sigma \, {\rm part}}^{\rm dec}=f\rho_{\phi}^{\rm dec}$ will initially be relativistic, becoming non-relativistic after an amount of expansion ${m_{\phi}/m_{\sigma}}$. Putting all this together, the density of decay curvatons at NR is given by
 \begin{equation}
 \rho^{\rm NR}_{\sigma \, {\rm part}} = f \left(\frac{m_\sigma}{m_\phi}\right)^4 \rho_\phi^{\rm dec} \,.
 \end{equation}

We have no need to evolve the densities further forward in time, since we can see from Fig.~\ref{sequence} that the late-time value of $r_{\rm dec}$, i.e.\ $r_{\rm dec}^{\rm max}$, is simply given by using Eq.~(\ref{e:rdec}) ignoring the contribution from radiation (as it disappears at late times, the ratio of the remaining terms being unchanged).

To compute the curvaton condensate density at NR, starting with its initial value $m_\sigma^2 \sigma_*^2/2$, as discussed above there are two parameter regions of relevance.  Again we track through relevant matter and radiation-dominated periods assuming instantaneous transitions, finding
\begin{eqnarray}
\rho_\sigma^{\rm NR} = \frac{1}{2} \sigma_*^2 \frac{m_\sigma^3}{m_\phi^3} \Gamma_\phi^2 &\quad ; \quad & m_{\sigma} > \Gamma_{\phi} \,;\\
\rho_\sigma^{\rm NR} = \frac{1}{2} \sigma_*^2 \frac{m_\sigma^{7/2}}{m_\phi^3}  \Gamma_\phi^{3/2} &\quad ; \quad & m_{\sigma}< \Gamma_{\phi} \,.
\end{eqnarray}
These of course match smoothly over the boundary \mbox{$m_\sigma = \Gamma_\phi$} where the two events happen simultaneously.\footnote{Strictly speaking these expressions are valid only if the curvaton condensate is sub-dominant to the radiation up to NR, but if this were not the case $r_{\rm dec}^{\rm max}$ will be extremely close to one and can be assumed to be one.}

\begin{figure*}[t]
\begin{center}
\vspace*{-1cm}
\includegraphics[width= 0.48\textwidth]{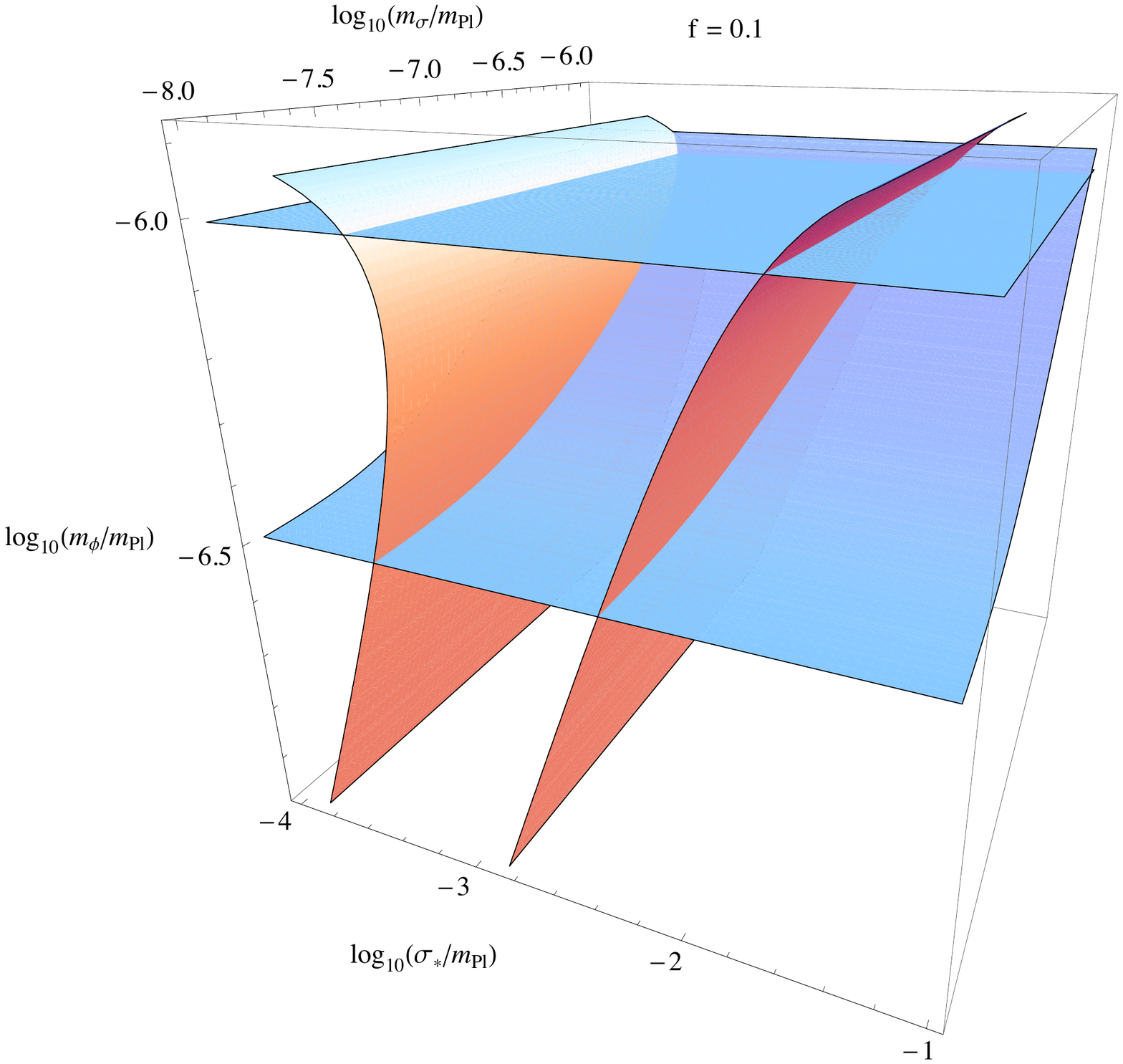}
\includegraphics[width= 0.48\textwidth]{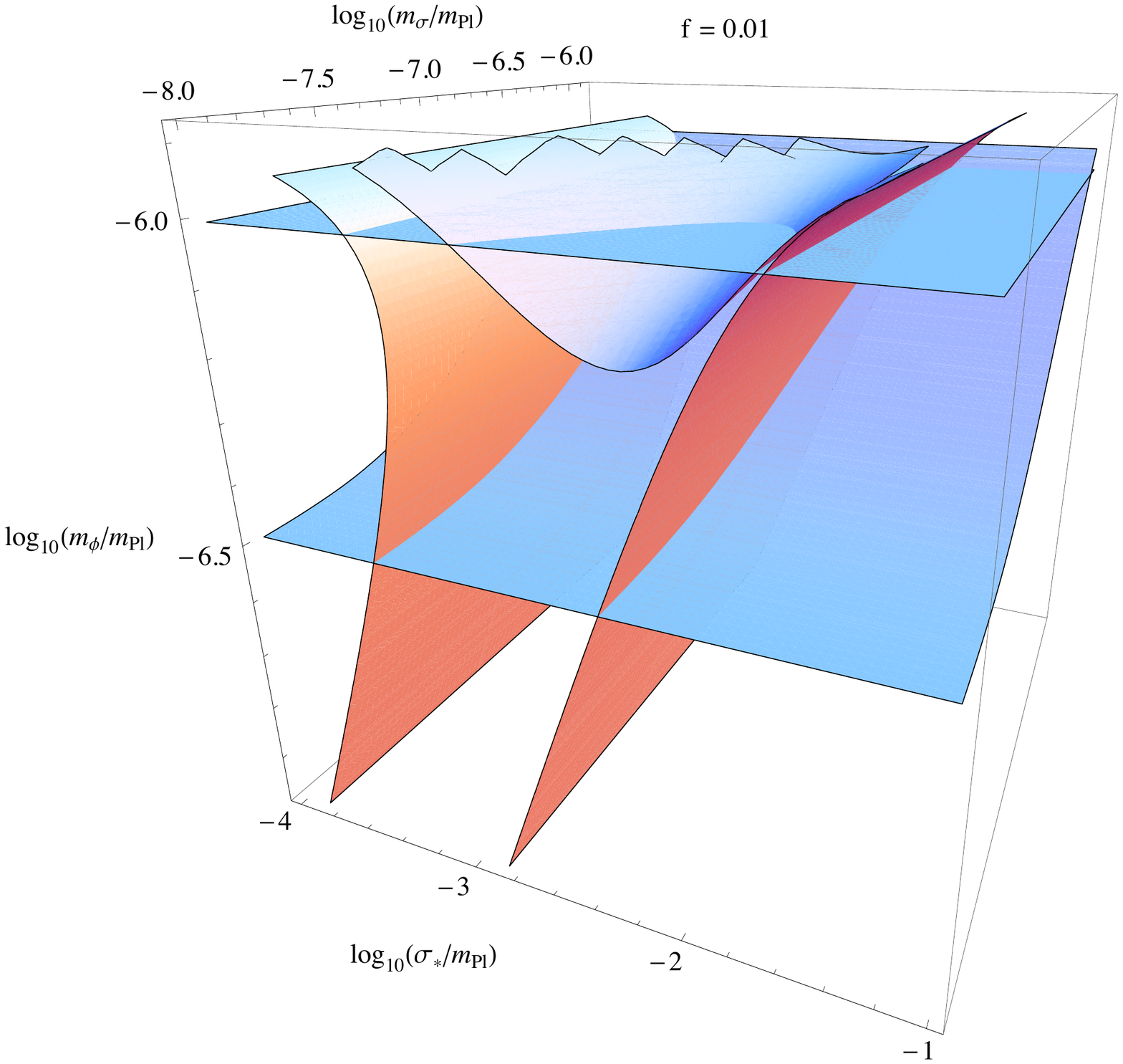}\\
\vspace*{-2.5cm}
\includegraphics[width= 0.48\textwidth]{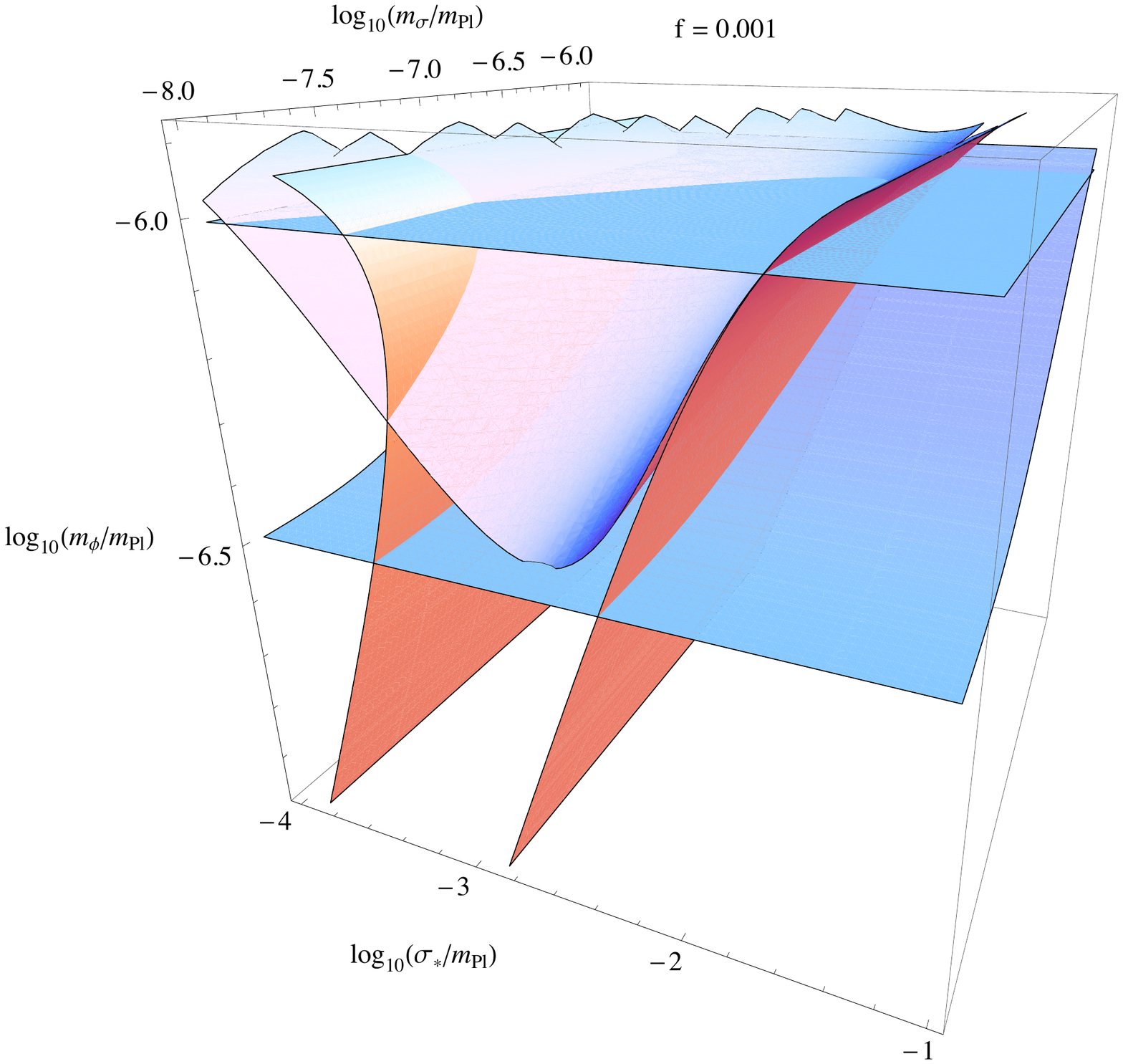}
\includegraphics[width= 0.48\textwidth]{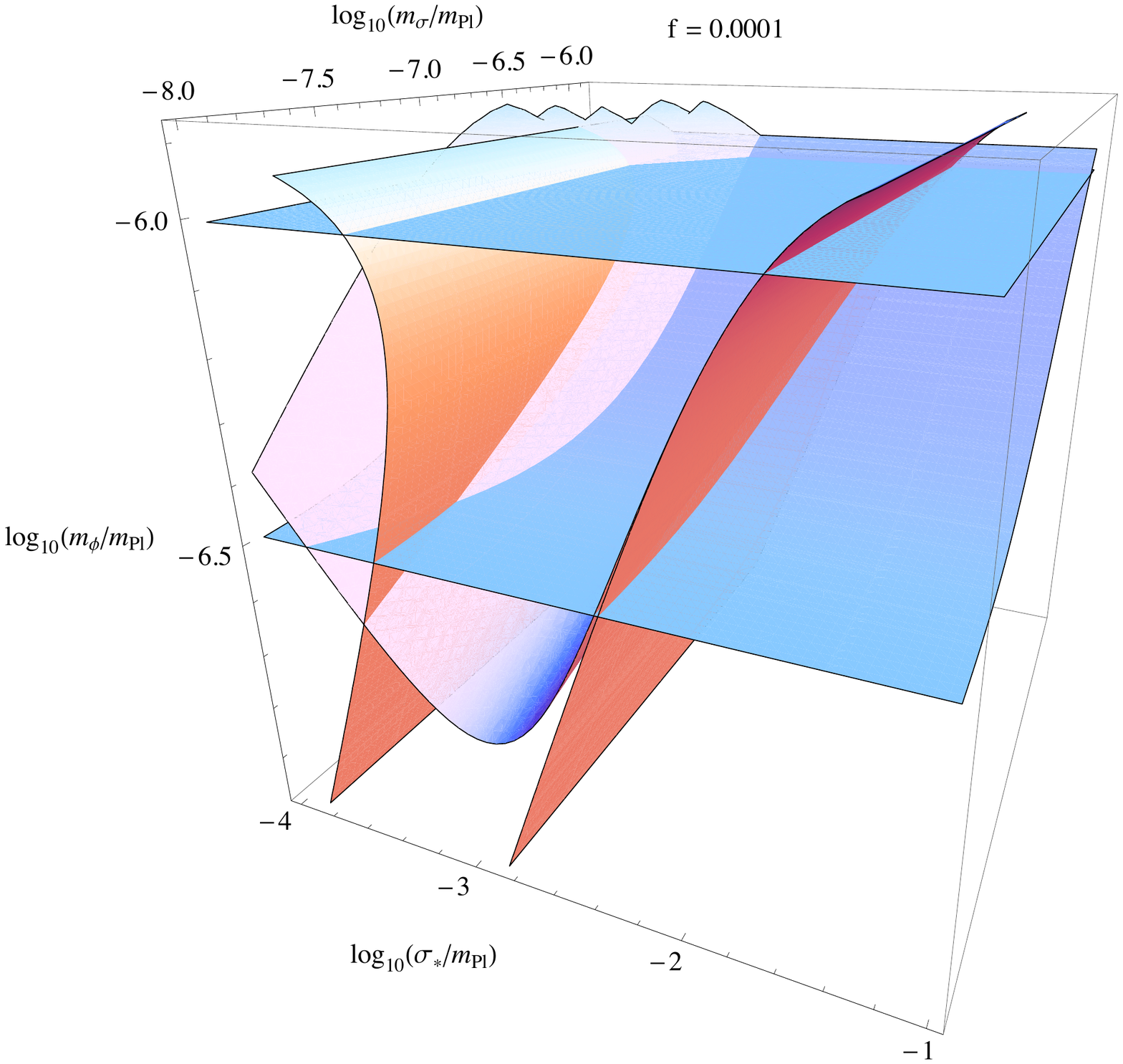}
\vspace*{-2cm}
\end{center}
\caption{Constraint plots for four different values of $f$, namely 0.1, 0.01, 0.001, and 0.0001. The original four constraint surfaces are as in Fig.~\ref{fzerocheese}. The new scoop-shaped surface (not present in the top-left plot as it is off the top of the region viewed) delineates the region where the decay curvatons prevent $r_{\rm dec}$ from reaching the value required to generate the observed amplitude of perturbations, i.e.\ $r_{\rm dec}^{\rm max} < r_{\rm dec}^{\rm observed}$. The allowed region is above the scoop. This constraint surface replaces the previous right-hand surface, which we still show for comparison.}
\label{fcheese} 
\end{figure*}

Finally, then
\begin{equation}
r_{\rm dec}^{\rm max} = \frac{1}{1+ \rho_{\sigma \, {\rm part}}^{\rm NR}/\rho_\sigma^{\rm NR}}\,,
\end{equation}
which enables us to compute $r_{\rm dec}^{\rm max}$ as a function of our base parameters $m_\phi$, $m_\sigma$, and $\sigma_*$ along with the new branching ratio parameter $f$.

\subsection{Constraints}

Figure~\ref{fcheese} shows the constraints for various values of the branching ratio. The new surface is the one where $r_{\rm dec}^{\rm max}$ is equal to the observed value of $r_{\rm dec}$; models below this surface cannot achieve a big enough $r_{\rm dec}$ to give the required perturbation amplitude and are hence excluded. This surface replaces the previous $r_{\rm dec} < 1$ surface (which we still include in the plots for comparison, though the new constraint is always stronger by construction).

First, we see by comparison to Fig.~\ref{fzerocheese} that the observational limits coming from $n_{\rm s}$, $r$ and $\fnl$ are unchanged by the introduction of $f$, as advertised, i.e.\ those surfaces are in the same position as before.

Second, we see that the decay curvatons can have a substantial effect on viability of the scenario if the branching ratio is high enough. For $f=0.1$, which admittedly is implausibly high, the entire displayed parameter regime gets excluded (though an allowed region reappears if the $m_\sigma$ axis is extended to much smaller values, necessitating a large hierarchy between the inflaton and curvaton masses). For a more reasonable value of $f = 0.01$, the new constraint bites off a substantial part of the previously-allowed parameter space, while for $f= 0.001$ the effect is more limited but still present. By the time $f$ reaches 0.0001, the new constraint no longer impacts on the region allowed by other constraints.

Hence we can conclude that a branching ratio of \mbox{$f < 10^{-4}$} is essentially irrelevant to the curvaton scenario and can be ignored, whereas branching ratios at the level of $f \gtrsim 0.1$ are incompatible with the simplest curvaton model unless there is a large mass hierarchy between inflaton and curvaton. Between those limits, whether the branching ratio is allowed or not depends on what values are assumed for the other parameters within their allowed ranges.
 
\section{Conclusions}\label{sec:conclusions}

In this article we assessed the impact on observables of the decay of the inflaton into curvaton particles during reheating. In particular we studied whether this channel of inflaton-to-curvaton decay might alleviate any source of tension between the curvaton scenario and observations, especially the tension with models that predict large $\fnl$. The answer is `no'. All that is relevant to the observables is the relative density of the inflaton decay products at the time of curvaton decay. Apart from this, it doesn't matter what route the inflaton decays through to reach the radiation bath; its adiabatic perturbations are super-horizon and won't change.

The crucial ingredient to uncover this physics is to make the correct generalization of $r_{\rm dec}$, the fractional curvaton density when it decays, in the presence of the decay curvatons.  It is very tempting to include the decay curvatons in both the numerator (the curvaton density) and the denominator (the total density), as has indeed been done in previous papers \cite{Sasaki:2006kq,Ade:2015ava}. We have argued that since the decay curvatons carry only inflaton-like perturbations into the final thermal bath, they should not be included in the curvaton density, only the total density. Perhaps the easiest way to see this is to imagine that the decays were instead into a different particle species with exactly the same properties as the curvaton particles; clearly the physical outcome will be completely the same, and there would be no reason to include those particles in the numerator. We stress that the results in Ref.~\cite{Sasaki:2006kq} are still correct, but conceal the degeneracy between their definition of $r_{\rm dec}$ and $\Delta_s^2$ that ensures all physical quantities are unchanged.

We saw that rather than helping the models, considering inflaton-to-curvaton decays might worsen the fit to data by preventing the curvaton condensate dominating the background dynamics. For this reason we have taken the relevant quantity to be $r_{\rm rdec}^{\rm max}$, which is the maximum achievable $r_{\rm dec}$ for given values of the free model parameters $m_{\phi}$, $m_{\sigma}$, and $\sigma_*$ and the branching ratio $f$. Using $r_{\rm rdec}^{\rm max}$ saturates the flexibility of the model to compensate for the lowering of $r_{\rm dec}$ caused by introducing $f$, and as such presents the best chance for the model to survive the introduction of the branching ratio. On the other hand, because $r_{\rm dec}^{\rm max}$ gives the most optimistic scenario for the curvaton it will give conservative constraints on $f$; stronger constraints might be obtained if the arbitrarily-late curvaton decays were not permitted.

For given values of the base parameters $m_\phi$, $m_\sigma$, and $\sigma_*$, any model that fits the data with $f \neq 0$ is degenerate with a model with $f=0$ and a different value of $\Gamma_\sigma$, as far as cosmological observations are concerned. Hence observations can place limits on $f$ (as a function of other model parameters) but cannot demonstrate a need for it to be non-zero. Depending on the range of parameters considered the constraints on $f$ vary from $10^{-4}$ to 1, with that highest value allowed only when $m_\sigma \ll m_\phi$. Allowing for a greater hierarchy between the inflaton and curvaton masses relaxes the constraints on $f$.

Finally, we note that while we have focussed on inflaton-to-curvaton decays, our qualitative results hold more generally for other routings of the inflaton energy density to the final radiation bath. This includes temporary matter domination by a massive decay product other than the curvaton, or even a subsequent inflationary epoch such as a bout of thermal inflation \cite{Lyth:1995ka}. The over-riding point that the final form of perturbations depends only on the relative mix of inflaton-originated and curvaton-originated radiation in the final bath remains true. Hence matching the observed perturbation amplitude continues to ensure all other observables are unchanged.

\begin{acknowledgments}
C.B.~was supported by a Royal Society University Research Fellowship, M.C.\ by EU FP7 grant PIIF-GA-2011-300606 and FCT (Portugal) grant reference SFRH/BPD/111010/2015, and A.R.L.\  by the Science and Technology Facilities Council [grant number ST/K006606/1]. Research at Perimeter Institute is supported by the Government of Canada through Industry Canada and by the Province of Ontario through the Ministry of Research and Innovation. We thank Andrei Linde for raising the issue of inflaton-to-curvaton decays that inspired this article, and Peter Adshead, Takeshi Kobayashi, Marco Peloso, Jussi Valiviita, and David Wands for discussions.
\end{acknowledgments}


\end{document}